\documentclass[11pt,letterpaper]{JHEP}

\usepackage{amsmath,amssymb,epsfig}
\usepackage[all]{xy}

\renewcommand{\a}{\alpha}
\renewcommand{\b}{\beta}

\renewcommand{\d}{\delta}

\newcommand{\z}{\psi}
\renewcommand{\k}{\kappa}

\newcommand{\m}{\mu}

\newcommand{\p}{\pi}

\newcommand{\s}{\sigma}
\renewcommand{\t}{\tau}

\renewcommand{\z}{\zeta}

\newcommand{\D}{\Delta}

\renewcommand{\S}{\Sigma}

\newcommand{\ZZ}{\mathbb{Z}}

\newcommand{\CC}{\mathbb{C}}
\newcommand{\PP}{\mathbb{P}}

\newcommand{\bbz}{\mathbb{Z}}

\newcommand{\bbq}{\mathbb{Q}}
\newcommand{\cp}[1]{{\mathbb{P}^{#1}}}
\newcommand{\tP}{\t_{\cp{1}}}

\newcommand{\cO}{\mathcal{O}}
\newcommand{\cL}{\mathcal{L}}
\newcommand{\cN}{\mathcal{N}}

\newcommand{\tW}{\widetilde{W}}
\newcommand{\tN}{\widetilde{\cN}}

\DeclareMathOperator{\id}{id}
\DeclareMathOperator{\rk}{rk}
\DeclareMathOperator{\ch}{ch}
\DeclareMathOperator{\Pic}{Pic}
\DeclareMathOperator{\Span}{Span}

\newcommand{\pt}{\textrm{pt}}

\newcommand{\FM}{\textbf{\textit{FM}}}
\newcommand{\T}{\boldsymbol{T}}

\newcommand{\Hecke}{\textbf{\textit{Hecke}}}


\title{Standard-Model Bundles on Non-Simply Connected Calabi--Yau
   Threefolds}

\author{Ron Donagi\\
   Department of Mathematics, University of Pennsylvania,
   Philadelphia, PA 19104--6395, USA.}
\author{Burt A.~Ovrut\\
   Department of Physics, University of Pennsylvania, Philadelphia, PA
   19104--6396, USA.} 
\author{Tony Pantev\\
   Department of Mathematics, University of Pennsylvania,
   Philadelphia, PA 19104--6395, USA.}
\author{Daniel Waldram\\
   Theory Division, CERN CH-1211, Geveva 23, Switzerland.\\
   Department of Physics, The Rockefeller University, 1230 York
   Avenue, New York, NY 10021.}

\abstract{We give a proof of the existence of $G=SU(5)$, stable
  holomorphic vector bundles on elliptically fibered Calabi--Yau
  threefolds with fundamental group $\bbz_2$. The bundles we construct
  have Euler characteristic $3$ and an anomaly  that can be absorbed
  by M-theory five-branes. Such bundles provide the
  basis for constructing the standard model in heterotic
  M-theory. They are also applicable to vacua of the weakly coupled
  heterotic string. We explicitly present a class of three family models
  with gauge group $SU(3)_C\times SU(2)_L\times U(1)_Y$.}

\preprint{UPR-893T, CERN-TH/2000-202, RU-00-4B}

\begin{document}


\section{Introduction}
\label{sec:intro}

Ho\v rava--Witten theory~\cite{hw} can be consistently compactified on a
Calabi--Yau threefold with non-vanishing four-form
``$G$-flux''~\cite{w}. It was shown in~\cite{losw} that this reduction
leads to a new low-energy limit of M-theory. This limit consists of a
five-dimensional ``bulk'' space with $N=1$ local supersymmetry,
bounded by two, four-dimensional, $N=1$ supersymmetric
$\ZZ_{2}$-orbifold fixed planes. Furthermore, the theory can admit BPS
fivebranes located in the bulk space, each with two spacelike
directions wrapped on a holomorphic curve in the Calabi--Yau
threefold~\cite{w,nse}. The worldvolumes of these wrapped fivebranes
possess $N=1$ supersymmetry. Hence, this limit of M-theory, called
heterotic M-theory, provides an explicit description of a ``brane''
universe, derived directly from a fundamental theory.

In addition to its brane structure, heterotic M-theory can also
account for much of phenomenological particle physics. In
keeping with the brane context, three families of $N=1$ supersymmetric
quarks and leptons can be shown to exist naturally on one of the
orbifold fixed planes, which we call the ``observable'' brane. These
transform under either grand unified gauge groups, such as $SO(10)$ and
$SU(5)$, or under the standard model gauge group, 
$SU(3)_{C} \times SU(2)_{L} \times U(1)_{Y}$. The supersymmetric
matter and the associated gauge groups arise from the higher
dimensional Ho\v rava--Witten theory as follows. Ho\v rava--Witten
theory consists of an eleven-dimensional $N=1$ locally supersymmetric
bulk space bounded by two ten-dimensional $\ZZ_{2}$-orbifold
fixed planes exhibiting $N=1$ supersymmetry. It was shown in~\cite{hw}
that anomaly cancellation requires the existence of an $N=1$, $E_{8}$
Yang--Mills supermultiplet on each of the orbifold fixed planes. 
This theory is then dimensionally reduced on a Calabi--Yau threefold,
$Z$, with a non-vanishing four-form flux as required by anomaly
cancellation. On either one of the orbifold fixed planes, the result
at low energy is a  four-dimensional, $N=1$ supersymmetric theory
with matter and gauge group content arising from the ``decomposition''
of the $E_{8}$ supermultiplet under the dimensional reduction. This
decomposition is entirely controlled by the vacuum structure of the
$E_{8}$ gauge fields on the Calabi--Yau threefold. As discussed
in~\cite{nse}, some subgroup $G \subseteq E_{8}$ of the gauge fields
can be non-vanishing on $Z$. These ``$G$-instantons'' will preserve
$N=1$ supersymmetry as long as they satisfy the Hermitian Yang--Mills
equations. However, the low energy gauge group is altered by these
instantons, being spontaneously broken from $E_{8}$ to $H$, where $H
\subseteq E_{8}$ is the commutant of the instanton structure group
$G$, that is, the largest subgroup in $E_8$ such that $[H,G]=0$. This
mechanism, originally stated in~\cite{hos}, 
allows GUT groups, such as $SO(10)$ and $SU(5)$ (the commutants of
$G=SU(4)$ and $SU(5)$ respectively), and the standard model gauge
group (the commutant of $G=SU(5) \times \bbz_{2}$, for example) to appear
on the observable brane. Furthermore, the decomposition of the $E_{8}$
Yang--Mills supermultiplet under $G$ into $H$ multiplets determines
the structure of matter on the observable brane. 

The construction of heterotic M-theory models with grand unified
gauge groups $H$ and three families of matter is relatively
straightforward, and has been discussed 
in~\cite{dlow1,dlow2,moduli,curio,andreas}. Specifically, such
theories require the construction of $G$-instantons on a simply
connected elliptically fibered Calabi--Yau threefold with a zero
section. Such instantons can be computed using the theorem of
Donaldson~\cite{donaldson} and Uhlenbeck--Yau~\cite{UhlYau}, which
relates them to polystable holomorphic vector bundles on $Z$, and
extensions~\cite{dlow1,dlow2} of the work of Friedman, Morgan and
Witten~\cite{fmw}, Donagi~\cite{donagi} and Bershadsky~\textit{et
  al.}~\cite{bjps}, which constructs vector bundles via the method of
spectral covers. However, constructing $G$-instantons that lead to the
standard model gauge group $H=SU(3)_{C} \times SU(2)_{L} \times
U(1)_{Y}$ and matter content is considerably more difficult. The
reason is that, in addition to vector bundles with continuous
structure group $G$, one must introduce Wilson lines into the
theory~\cite{candelas,wilsonline}. The existence of Wilson lines, however,
requires that the Calabi--Yau threefold $Z$ have non-trivial
fundamental group. Such manifolds do not admit a zero section and the
vector bundle construction of~\cite{fmw,donagi,bjps} no longer applies.   

To overcome this problem, it is necessary to give a method for
computing polystable, holomorphic vector bundles over a
non-simply connected Calabi--Yau threefold. A major step in this
direction was taken in~\cite{z2}, where a spectral cover formalism
applicable to torus fibrations without zero section was
presented. (Similar constructions were considered in~\cite{ack}). The 
torus fibrations, $Z$, were constructed as a quotient, $Z=X/\t_X$,
where $X$ is a simply connected Calabi--Yau threefold with two
sections and $\t_X$ is a freely acting involution interchanging the
sections.  Finding phenomenologically acceptable heterotic M-theory
vacua  in this context amounts to finding $\t_{X}$-invariant stable vector
bundles on $X$ satisfying certain conditions (anomaly cancellation and
3-families). These conditions involve
only the charges of the corresponding vacua, which mathematically are
encoded in the Chern classes. The $\tau_{X}$-invariance imposes
non-trivial restrictions on the charges. In \cite{z2} we exhibited an
infinite collection of $\tau_{X}$-invariant admissible charges and
showed that each of those is realized by a non-empty family of vector
bundles on $X$. The next step is to prove the existence of actual
$\tau_{X}$-invariant vacua in such families. This is the subject of
the present work.

Such a proof is considerably more difficult than showing the
invariance of the Chern classes alone. Rather than consider the
bundles presented in~\cite{z2}, for various technical reasons, it is
expedient to consider a different class of vector bundles over a 
Calabi--Yau threefold with $\p_1(Z)=\ZZ_{2}$. These bundles also lead
to the standard model, but have a structure that lends itself more
easily to discussing invariance under $\ZZ_{2}$. In this paper, we
explicitly present polystable, holomorphic vector bundles of this
type and prove that they are $\ZZ_{2}$-invariant. This is a
fundamental step in demonstrating that the standard model can arise in
heterotic M-theory. This construction, and the proof of the $\ZZ_{2}$
invariance, is necessarily of a technical nature but is sufficiently
important for phenomenological physics that we synopsize it in this
paper. The complete proof, with all mathematical details, is
presented in a companion papers~\cite{math}. 

The types of invariant bundles discussed here can be extended
to larger freely acting automorphism groups of $X$, such as $\bbz_3$,
$\bbz_2\times\bbz_2$ and so on. We will show in a future paper 
that standard-like models arising from $SO(10)$ and other grand unified
groups can be constructed with the possibility of suppressed nucleon
decay. Finally, we would like to emphasize that, although the vector
bundles discussed in~\cite{dlow1,dlow2,z2} and this paper are within
the context of heterotic M-theory, these bundles are equally
applicable to the construction of new, phenomenologically interesting
vacua in the weakly coupled heterotic string.


\section{Outline of the Construction}
\label{sec:outline}

Let us start by summarizing the problem we have to solve. Recall from
the introduction that to construct an $N=1$ vacuum in four-dimensions,
we must compactify on a Calabi--Yau threefold $Z$ and choose $E_8$
gauge fields in a $G\subseteq E_8$ polystable holomorphic vector bundle
$\mathcal{V}$ on $Z$. 

We would like to construct models with the standard model gauge group
and three families of charged matter. In this paper, we will break the
gauge group through an $SU(5)$ GUT group. We have
\begin{equation}
   E_8 \stackrel{\text{bundle }\mathcal{V}}{\longrightarrow}
      SU(5) \stackrel{\bbz_2\text{ Wilson line}}{\longrightarrow}
      SU(3)_C \times SU(2)_L \times U(1)_Y . 
\label{eq:breaking}
\end{equation}
By choosing an $SU(5)$ vector bundle $\mathcal{V}$ on the Calabi--Yau
manifold $Z$, we break the preserved gauge group to the commutant
$SU(5)$. By choosing a $\bbz_2$ Wilson line, we can then further break
this down to the standard model gauge group $SU(3)_C \times SU(2)_L
\times U(1)_Y$.

We recall that there are two additional
conditions~\cite{dlow1,dlow2}. First, the requirement of three
families translates into a condition on the third Chern class
\begin{equation}
   N_{\text{gen}} = \frac{1}{2}c_3(\mathcal{V}) = 3 . 
\label{eq:three}
\end{equation}
Second, in order to cancel anomalies the orbifold planes and the
fivebranes must be sources of $G$-flux. The condition that the net charge
vanishes becomes
\begin{equation}
   c_2(\mathcal{V}) + [W] = c_2(TZ) ,
\label{eq:eff}
\end{equation}
where $[W]$ is the total cohomology class of the holomorphic curves on
which the fivebranes are wrapped. To describe physical branes,
the class $[W]$ must be effective.

In general, then, we need to satisfy the following conditions 
\begin{itemize}
\item[]\hspace{-4ex}\textbf{Supersymmetry:} $Z$ is a Calabi--Yau
   threefold. $\mathcal{V}$ is a polystable, holomorphic $SU(5)$ bundle, 
\item[]\hspace{-4ex}\textbf{Wilson Line:} $Z$ has, at least,
   $\p_1(Z)=\bbz_2$,
\item[]\hspace{-4ex}\textbf{Anomaly Cancellation:} $c_2(TZ) -
   c_2(\mathcal{V})$ must be an effective class in $Z$,
\item[]\hspace{-4ex}\textbf{Three Families:} $c_3(\mathcal{V})=6$, 
\end{itemize}
in order to have a realistic model. 

The simplest way to construct a suitable Calabi--Yau threefold $Z$ is as a
quotient~\cite{candelas}. Let $X$ be a smooth Calabi--Yau threefold
with trivial fundamental group. Suppose, in addition, we have an
involution, 
\begin{equation}
   \t_X : X \to X ,
\label{eq:tXdef}
\end{equation}
with $\t_X^2=\id$, which is freely acting, that is, has no fixed
points, and which preserves the holomorphic 3-form. 
The quotient space $Z$ formed by identifying points related by
the involution,
\begin{equation}
   Z = X/\t_X , 
\label{eq:Z}
\end{equation}
is then a smooth Calabi--Yau threefold with $\p_1(Z)=\bbz_2$. 

Rather than construct the bundle $\mathcal{V}$ directly on the quotient
$Z=X/\t_X$, it is generally easier to construct it from a bundle $V$ on
$X$. If $V$ is $\t_X$-invariant, that is 
\begin{equation}
   \t_X^*V \cong V ,
\label{eq:Vinv}
\end{equation}
then, as long as $V$ is stable,  it will descend 
to a bundle $\mathcal{V}$ on $Z$. 

We will construct the bundle $V$ via the ``spectral cover''
construction~\cite{fmw,donagi,bjps} which, essentially, uses T-duality
to describe $V$ in terms of simpler T-dual data. However, this puts a
constraint on the Calabi--Yau manifold $X$. It requires that $X$ is
elliptically fibered over a two-complex-dimensional base. This means
that at each point on the base, there is a torus fiber on which one
can perform the T-duality. We should note also that, in general, the
spectral construction gives a $U(n)$ rather than an $SU(n)$
bundle. Thus, we will need the additional condition that $c_1(V)=0$ in
order to ensure that $V$ is an $SU(5)$ bundle. 

In summary then, if $Z$ is a quotient manifold and we build the bundle via the
spectral construction, we must satisfy the following conditions 
\begin{quote}
\begin{enumerate}
\renewcommand{\labelenumi}{\textbf{(Z2)}}
\renewcommand{\theenumi}{\textbf{(Z2)}}
\item \label{Z2}
   $X$ is a smooth elliptically fibered Calabi--Yau threefold
   admitting a freely-acting involution $\t_X:X\to X$, 
\renewcommand{\labelenumi}{\textbf{(S)}}
\renewcommand{\theenumi}{\textbf{(S)}}
\item \label{S}
   $V$ is a stable rank-five vector bundle on $X$,
\renewcommand{\labelenumi}{\textbf{(I)}}
\renewcommand{\theenumi}{\textbf{(I)}}
\item \label{I}
   $V$ is $\tau_{X}$-invariant,
\renewcommand{\labelenumi}{\textbf{(C1)}}
\renewcommand{\theenumi}{\textbf{(C1)}}
\item \label{C1}
   $c_1(V) = 0$, 
\renewcommand{\labelenumi}{\textbf{(C2)}}
\renewcommand{\theenumi}{\textbf{(C2)}}
\item \label{C2}
   $c_2(TX) - c_2(V)$ is effective,
\renewcommand{\labelenumi}{\textbf{(C3)}}
\renewcommand{\theenumi}{\textbf{(C3)}}
\item \label{C3}
   $c_3(V) = 12$ .
\end{enumerate}
\end{quote}
Note that the final three-family condition is now $c_3(V)=12$ because,
under the quotient, we have $c_3(\mathcal{V})=\frac{1}{2}c_3(V)$. 

Our problem, then, is to find solutions to the
conditions~\ref{Z2}--\ref{C3}. The procedure, which is summarized in
Figure~\ref{fig:flow}, will be as follows. 
First, in Section~\ref{sec:XtX}, we construct a large family of
elliptically fibered Calabi--Yau manifolds $X$ satisfying the involution
condition~\ref{Z2}. Second, in Section~\ref{sec:V}, we construct a
large family of bundles $V$ on $X$ satisfying the invariance
condition~\ref{I}. This is the most difficult part of the
construction. Finally, in Section~\ref{sec:num}, we reduce the
stability condition~\ref{S} and the conditions on the Chern
classes~\ref{C1}--\ref{C3} to numerical conditions on the parameters
defining $V$. A class of solutions to these conditions is then given
in Section~\ref{sec:examples}. 
\FIGURE{%
   \xymatrix@R=1pc{
   *\txt{\framebox{\framebox{\parbox{2.5in}{
       Construct a large family of elliptically fibered Calabi--Yau
       threefolds $X$ with a  freely acting involution $\tau_X$ to
       satisfy the condition~\ref{Z2} 
       \hfill (\ref{sec:XtX}) }}}} \ar[d] & 
   *\txt{\framebox{\parbox{2.5in}{
       1. Construct a five-dimensional class of rational elliptic
       surfaces $B$ with an involution  $\tau_B$.
       \hfill (\ref{sec:B}),~(\ref{sec:tB}) \\ 
       2. Specialize to a class of $(B,\tau_B)$ with split fibers.
       \hfill (\ref{sec:specialB}) \\ 
       3. Construct $(X,\tau_X)$ as fiber product $B\times_{\cp{1}}B'$
       of special rational elliptic surfaces.
       \hfill (\ref{sec:X}) }}} \ar[l] \\ 
   *\txt{\framebox{\framebox{\parbox{2.5in}{
       Construct a family of $\tau_X$-invariant vector bundles $V$ for
       each $(X,\tau_X)$ via the spectral construction, thus satisfying the
       condition~\ref{I} 
       \hfill (\ref{sec:V}) }}}} \ar[d] &
   *\txt{\framebox{\parbox{2.5in}{
       4. Construct a class of $\tau_X$-invariant, rank-$r$ bundles
       $V_r$ as pullbacks of $\tau_B$-invariant bundles on $B$, 
       modified by Hecke transforms . 
       \hfill (\ref{sec:Vr}) \\ 
       5. Construct $V$ as an extension of $V_2$ and $V_3$ bundles.
       \hfill (\ref{sec:ext}) }}} \ar[l] \\
   *\txt{\framebox{\framebox{\parbox{2.5in}{
       Impose the conditions~\ref{S}, \ref{C1}, \ref{C2} and \ref{C3},
       which become a set of numerical conditions on the parameters
       defining $V$. 
       \hfill (\ref{sec:num}),~(\ref{sec:examples}) }}}}
   }
   \caption{Flow diagram of the construction. The numbers refer to
     the relevant section.}
   \label{fig:flow}
}

In order to simplify the construction of $\t_X$-invariant bundles,
we specialize to Calabi--Yau manifolds $X$ built from a particular
type of complex two-fold base $B$ known as a ``rational elliptic
surface'' or a ``$dP_{9}$''. These are surfaces which are themselves
elliptically 
fibered. In order to construct an involution $\t_X$ on $X$, we 
need to understand the action of involutions $\t_B$ on the surfaces
$B$. The structure of $B$ and the special class of $B$ admitting
$\t_B$, are discussed in Sections~\ref{sec:B} and~\ref{sec:tB}.

In order to find solutions to the numerical conditions, we must
actually specialize further and consider only those rational
elliptic surfaces where some of the elliptic fibers of $B$ split. 
This introduces
new effective classes on $B$ and, hence, new freedom in constructing
$V$. This four-dimensional sub-family of surfaces is described in
Section~\ref{sec:specialB}. The manifold $X$ is then constructed as
the ``fiber product'' of two surfaces $B$ and $B'$. (Note that
Calabi--Yau manifolds of this type were also considered in~\cite{ack}.)
The involution $\t_X$ is similarly built from involutions $\t_B$ and
$\t_B'$ on each of the two surfaces. This is described in
Section~\ref{sec:X}.

Turning to the construction of $V$, we may take as a first
approximation to $V$  a pullback to $X$ of a bundle $W$ on $B$. 
The point is that on
$B$, it is comparatively easy to find those bundles which are
invariant under the
corresponding involution $\t_B$, a problem which is harder to solve on
the full manifold $X$. However, bundles of this form are not general enough to
satisfy all the numerical conditions. One is forced to make two
generalizations. First, one modifies the bundle by means of ``Hecke
transforms'' which, roughly speaking, modify the bundle over surfaces in the
Calabi--Yau threefold. Here, it is the fact that the fibers of 
$B$ and $B'$ split
which provides new surfaces in $X$, and the additional freedom to make
Hecke transforms over these surfaces. This construction is described in
Section~\ref{sec:Vr}. 

Even with the additional freedom from Hecke transforms, the bundles one
constructs, $V_r$ for general rank $r$, cannot satisfy both~\ref{C1}
and~\ref{C3}. This leads to the second generalization discussed in
Section~\ref{sec:ext}: the final bundle $V$ is actually built as the
``extension'' of a rank-two bundle $V_2$ and a rank-three bundle $V_3$,
each built using the construction just described. This means that $V$
is a sort of ``twisted'' sum of $V_2$ and $V_3$. 

Finally, we show in Section~\ref{sec:examples} that there is then a
large class of solutions to the numerical conditions. For the example
we give, the solution has four independent parameters. Other classes
of examples also exist. Even given the constraints, we find that there
is still a large freedom in the examples.


\section{A Family of $(X,\t_X)$}
\label{sec:XtX}

Our first goal is to construct a large family of elliptically
fibered Calabi--Yau threefolds $\p:X\to B'$ with a freely acting
involution $\t_X$. Some explicit examples were described
in~\cite{z2}. Here, we will consider a more general class of
involutions but specialize to the case where the base $B'$ of the
fibration is a rational elliptic surface. This will lead
to a large class of examples for which the demonstration of the
invariance of the vector bundles is particularly simple.

Let us recall some properties of the construction
in~\cite{z2}. First, the involution $\t_X$ will necessarily induce
some involution $\t_{B'}$ of the base $B'$. Let $\D$ be the
discriminant of the fibration $\p$. Since $X$ is a Calabi--Yau
manifold, $\D$ is a section of $K_B^{-12}$. A necessary condition for
$\t_X$ to be freely acting is that the set of fixed points of
$\t_{B'}$ be disjoint from the discriminant locus of $\p$, that is  
\begin{equation}
   \{ \text{fixed points of $\t_B'$} \} \cap \{ \D=0 \} = \varnothing. 
\label{eq:tBcond}
\end{equation}
Consequently, the first step in the construction of $(X,\t_X)$ is to
find suitable base pairs $(B',\t_{B'})$. Much of this section will thus
concentrate on the construction of involutions on rational elliptic
surfaces.  

As discussed in section~\ref{sec:X}, the fact that the base $B'$ is
itself an elliptic fibration, $\b' : B' \to \cp{1}$, will allow us to
construct $X$ as the fiber product of a pair of rational elliptic
surfaces $B$ and $B'$ over a common $\cp{1}$. Thus, in this case,
simply by understanding involutions on rational elliptic surfaces we
will be able to construct involutions on $X$. Such constructions were
first considered in \cite{schoenCY}. In fact, exactly the four dimensional
subfamily of rational elliptic surfaces described in section
\ref{sec:specialB} below happened to appear, in a different context,
as an example in \cite[Section~9]{schoenCY}.

\subsection{Rational Elliptic Surfaces}
\label{sec:B}

A rational elliptic surface $B$ is a two-dimensional complex manifold
which is a fibration of elliptic curves over a sphere base $\cp{1}$ 
\begin{equation}
   \b : B \to \cp{1} .
\label{eq:res}
\end{equation}
It can be described as the blow-up of the projective plane $\cp{2}$ as
follows. Recall that an elliptic curve, which is topologically a
torus, corresponds to a cubic curve in $\cp{2}$. Consider a
one-parameter family or ``pencil'' of cubics where each curve passes
through nine fixed points $A_1,\dots,A_9$ in $\cp{2}$. The general
surface $B$ is then the blow-up of $\cp{2}$ at the nine points and the
elliptic fibers of $B$ just correspond to the cubic curves. Under mild
general position requirements, each subset of eight of the points
determines the pencil of cubics and, hence, the ninth point. This
implies that the rational elliptic surfaces depend on eight complex
parameters. (Fixing eight points in $\cp{2}$ requires 16
parameters. However, since only the relative positions matter, we must
subtract the dimension $\dim\PP\textrm{GL}(3,\CC)=8$ of the
automorphism group of $\cp{2}$, leaving eight parameters.) Although
not true del Pezzo surfaces, rational elliptic surfaces are sometimes
referred to as dP$_9$.  (A del Pezzo surface $dP_{n}$ is obtained by
blowing up $n \leq 8$ points in ${\mathbb P}_{2}$.)

Let $e_1,\dots,e_9$ be the classes of the exceptional divisors of $B$
corresponding to the nine blown-up points $A_i$. An additional divisor
$l$ comes from the class of a line in $\cp{2}$. It is easy to see that
these provide an independent basis for the cohomology of $B$, such
that 
\begin{equation}
   H^2(B,\bbz) = \bbz l \oplus 
      \left( \oplus_{i=1}^9 \bbz e_i \right) ,
\label{eq:Bcohomo}
\end{equation}
and, furthermore, the intersections between classes are given by
$l^2=1$, $l\cdot e_i=0$ and $e_i\cdot e_j=- \d_{ij}$. In addition, the
anti-canonical class of $B$ is equal to the class of the elliptic
fiber  
\begin{equation}
   -K_B = c_1(B) = 3l - \sum_{i=1}^9 e_i .
\label{eq:fB}
\end{equation}
which we denote by $f$.
Finally, we note that since each exceptional curve $e_1,\dots,e_9$
intersects the fiber at one point, these are all sections of the
fibration. In this paper, we will always identify one section
$e:\cp{1}\to B$ as the zero section. Without loss of generality we can
take $e=e_9$. Fixing the zero section determines a group law 
generating the translational symmetries of each smooth (toroidal)
fiber of $B$.

\subsection{Involutions of Rational Elliptic Surfaces}
\label{sec:tB}

Let us now consider involutions of $B$. Given the group action on the
fibers, one can define a natural involution $(-1)_B:B\to B$ of any
rational elliptic surface as the extension to all of $B$ of 
\begin{equation}
   (-1)_B(x) = -x , 
\label{eq:-1B}
\end{equation}
where $x$ is any point on a smooth fiber. This is the usual inversion
symmetry of the torus so that, on any smooth fiber, the action of
$(-1)_B$ will have four fixed points. In particular, one notes that this
involution leaves the zero section invariant 
\begin{equation}
   (-1)_B(e) = e .
\label{eq:-1Be}
\end{equation}
Recall the requirement~\eqref{eq:tBcond} that the fixed points of $\t_B$
be disjoint from the locus $\D=0$ of the discriminant, which is a
section of $K_B^{-12}=\cO_B(12f)$. Clearly, this locus intersects the
zero section 12 times, and, hence, is not disjoint from the set of fixed
points of $(-1)_B$. Thus, $(-1)_B$ is unsatisfactory for construction
of a freely acting $\t_X$. 

Instead, we must specialize the rational elliptic surface to a family
that admits additional involutions. Since $B$ is elliptically fibered,
our discussion will be much like the discussion in~\cite{z2} for
elliptically fibered threefolds, though with one generalization. 

First note that any involution $\t_B$ will induce an involution
$\tP:\cp{1}\to\cp{1}$ in the base. Assuming that $\tP$ does not act as
the identity\footnote{Note that there is a whole second family of
  rational elliptic surfaces with $\t_B$ built on $\tP=\id_\cp{1}$. We
  will not discuss them here, but these too could be used to find
  suitable $(X,\t_X)$ for constructing particle physics vacua, very
  much along the lines of this paper.},
any $\tP$ will have two fixed points, which we will denote as
$0,\infty\in\cp{1}$ and which uniquely determine the involution. Let
us fix a particular $\tP$. One can then show that any $\t_B$
satisfying $\tP\circ\b=\b\circ\t_B$ can be built out of a pair of
objects $(\a_B,\z)$. First, one needs an involution $\a_B$ which is a
lift of $\tP$ and so satisfies $\tP\circ\b=\b\circ\a_B$. We also
assume it leaves the zero section invariant 
\begin{equation}
   \a_B(e) = e .
\label{eq:aB}
\end{equation}
Second, one needs a section $\z$ of $\b$ satisfying 
\begin{equation}
   \a_B(\z) = (-1)_B(\z) .
\label{eq:zeta}
\end{equation}
The involution $\t_B$ is then given by
\begin{equation}
   \t_B = t_\z \circ \a_B ,
\label{eq:tB}
\end{equation}
where $t_\z$ is the translation of the elliptic fibers defined by the
section $\z$ 
\begin{equation}
   t_\z(x) = x + \z(p) ,
\label{eq:tz}
\end{equation}
where $x$ is any element of a smooth fiber $\p^{-1}(p)$ over a point
$p\in\cp{1}$. The conditions~\eqref{eq:aB} and~\eqref{eq:tB} are
required to ensure $\t_X^2=\id_B$. Note that in the particular case
where $\z=e$, then $\t_B=\a_B$. Also note that this construction 
is a generalization
of the involutions considered in~\cite{z2}. In that paper, we
required that $(-1)_B(\z)=\z$ so that~\eqref{eq:zeta}
became $\a_B(\z)=\z$. 

One finds~\cite{math} that there is a natural (and 
unique up to a twist by $(-1)_{B}$) 
way to construct $\a_B$ from the Weierstrass model of
$B$. However, the construction does not exist for all rational
elliptic surfaces $B$. Rather,
within the eight parameter family of rational elliptic surfaces there
is a five-dimensional sub-family of surfaces
 which admit $\a_B$. In addition, all surfaces in this sub-family 
also admit a non-trivial
section $\z$ satisfying~\eqref{eq:zeta}. Rather than discuss the
details of the construction, which can be found in~\cite{math}, let us
simply summarize the fixed point structure of $\a_B$ and $\t_B$. 

Since the fixed points of $\tP$ are at $0$ and $\infty$, any fixed
points of an involution of $B$ lifting $\tP$ must lie in the fibers
$f_0=\b^{-1}(0)$ and $f_\infty=\b^{-1}(\infty)$. First consider
$\a_B$. One can show that it acts as the identity on one of the
fibers, say  $f_0$
and as $(-1)$ on $f_\infty$. Thus, the whole fiber $f_0$ is fixed
pointwise under $\a_B$, as are four fixed points in $f_\infty$. This
is shown schematically in Figure~\ref{fig:Bfp}. It might appear odd
that this action treats $f_0$ and $f_\infty$ asymmetrically. The point
is that the Weierstrass model naturally defines two involutions
preserving the zero section, $\a_B$ and $\a_B\circ(-1)_B$. Under the
latter involution, the fixed point structure of $f_0$ and $f_\infty$
is reversed. 

Let us now turn to $\t_B$. First we note that the
condition~\eqref{eq:zeta} implies that on $f_0$ we have
$\z(0)=-\z(0)$ (that is, it is one of the four fixed points of the
$(-1)$ involution on $f_0$), while there is no such condition on
$f_\infty$. Provided $\z$ is not the zero section, translation by $\z$
will thus remove all the fixed points on $f_0$. The $\a_B$-fixed
points on $f_\infty$ are simply translated by $\z(\infty)/2$. Thus
$\t_B$ has four fixed points as shown in Figure~\ref{fig:Bfp}. 
\FIGURE{%
   \parbox{\textwidth}{%
   \begin{center}
      \epsfig{file=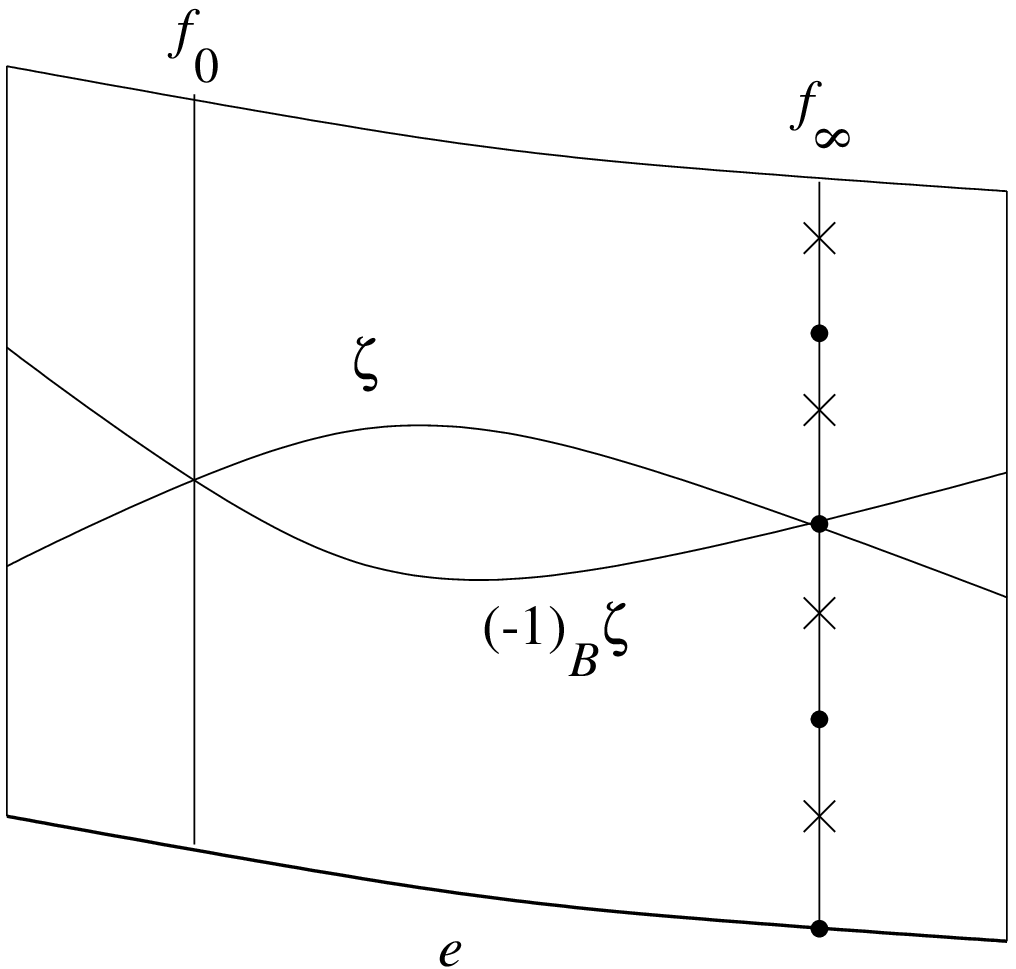,height=2.5in}
   \end{center}
   \caption{A general rational elliptic surface $B$ admitting $\a_B$.}
   \label{fig:Bfp}
}}

Clearly, in general, both $\a_B$ and $\t_B$ satisfy the
condition~\eqref{eq:tBcond}. However, as we will discuss in
section~\ref{sec:X} below, it is easy to show that only $\t_B$ leads
to a freely acting $\t_X$. Nonetheless, it would appear that we have
solved our problem, finding a five-dimensional family of $(B,\t_B)$
suitable for building the threefold $X$.

\subsection{Special Rational Elliptic Surfaces}
\label{sec:specialB}

It turns out that the sub-family of surfaces described in the previous
section is not quite suitable for our purposes. General rational
elliptic surfaces, including generic members of the sub-family, have
12 singular $I_1$ elliptic fibers where the torus pinches to a
sphere. In constructing $\t_X$-invariant bundles on $X$, we will find
that it is important for $B$ (and hence $X$) to have a richer
cohomology structure. In particular, we will require that $B$ has some
$I_2$ fibers where the torus splits into a pair of spheres. If the
split fiber is not to lie over a fixed point of $\tP$, we see that we
actually need at least one pair of $I_2$ fibers. These will lie above
a pair of points $p_1$ and $p_2$ in $\cp{1}$ exchanged by $\tP$. 

Thus the actual rational elliptic surfaces we will use in constructing
$X$ are a special four-dimensional sub-family of the family described
in the previous section, with, generically, a pair of $I_2$ fibers and 8
$I_1$ fibers. These surfaces admit $\a_B$ and $\t_B$ exactly as above,
and, in particular, the fixed points of $\t_B$ remain four points in
$f_\infty$ (see Figure~\ref{fig:special}).
\FIGURE{%
   \parbox{\textwidth}{%
   \begin{center}
      \epsfig{file=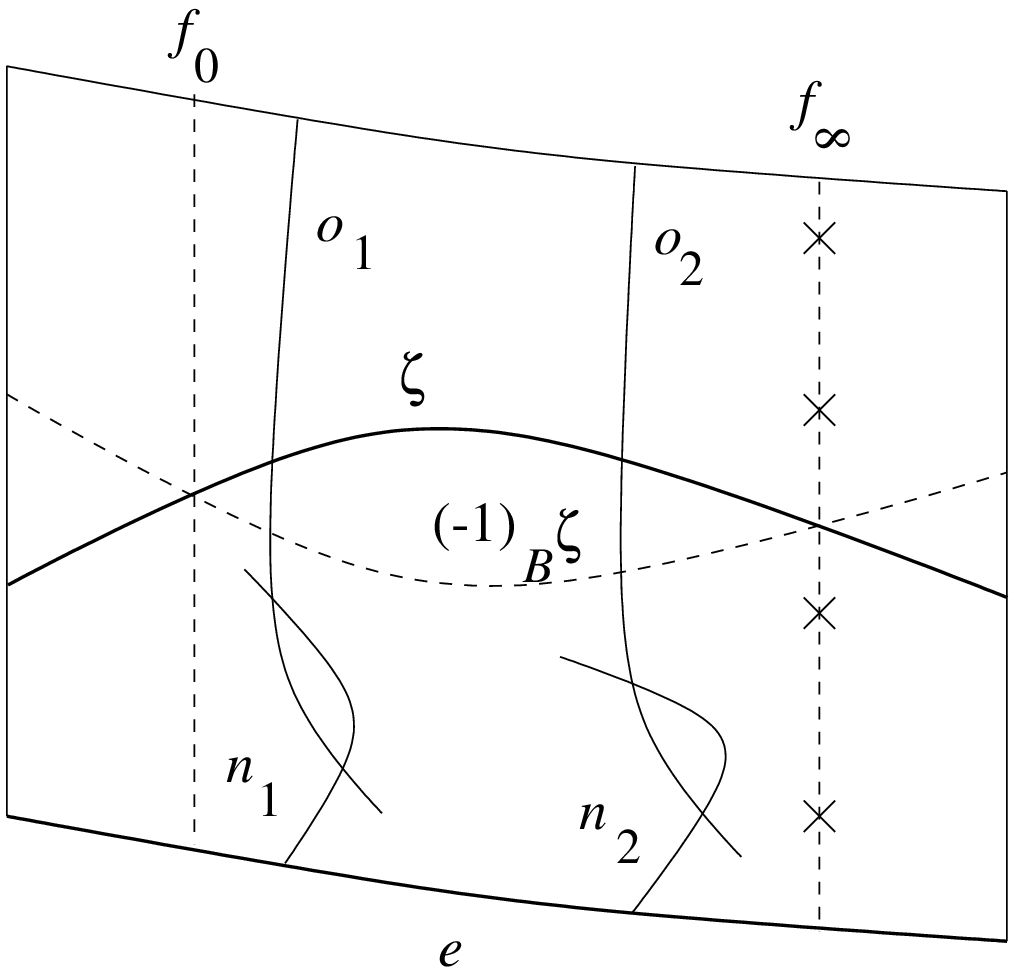,height=2.5in}
   \end{center}
   \caption{A special rational elliptic surface $B$.}
   \label{fig:special}
}}

Regarded as a blow-up of $\cp{2}$, the special features of this
sub-family translate into special position requirements on the nine
blow-up points. This is described explicitly in~\cite{math}. Again,
rather than discuss the details of the construction, let us simply
note the action of the $\t_B$ involution. We have already discussed the
fixed points of $\t_B$. What remains is to identify the cohomology
classes and the induced action of $\t_B$ on them. 
\TABLE{%
   \parbox{\textwidth}{%
   \begin{center}
   \begin{tabular}{|l||l|}
      \hline
                           & $\tau_{B}^{*}$ \\ 
      \hline 
        $e_1=\z$              & $e_9$ \\
        $e_j$ ($j=2,3$)    & $f - e_j + e_1 + e_9$ \\
        $e_i$ ($i=4,5,6$)  & $f - l + e_i + e_1 + e_7 + e_9$ \\
        $e_7$              & $l - e_2 - e_3$ \\
        $e_8$              & $f - l + e_1 + e_7 + e_8 + e_9$ \\
        $e_9=e$              & $e_1$ \\
        $l$                & $2f + 2(e_1+e_9) - (e_2+e_3) + e_7$ \\
      \hline
        $f$                & $f$   \label{tab:tB} \\
      \hline 
   \end{tabular}
   \end{center}
   \caption{Action of $\t_B^*$ on $H^2(B,\ZZ)$}
}}

Let the $I_2$ fibers be $f_1$ and $f_2$. Then each fiber is a union of
spheres 
\begin{equation}
   f_1 = n_1 \cup o_1 , \qquad
   f_2 = n_2 \cup o_2 . 
\label{eq:I2}
\end{equation}
As in the general case~\eqref{eq:Bcohomo}, the cohomology of $B$ can
be described in terms of nine exceptional divisors $e_i$ and the
pre-image $l$ of the line in $\cp{2}$. As usual, we will identify
$e=e_9$ as the zero section. In addition, we can identify $\z=e_1$ as
the section defining $\t_B$. The explicit construction then
identifies the new effective classes $n_i$ and $o_i$ as follows
\begin{equation}
\begin{aligned}
   n_1 &= e_8 - e_9 , \\
   o_1 &= 3l - e_1 - \dots - e_7 - 2e_8 = f - n_1 , \\
   n_2 &= l - e_7 - e_8 - e_9 , \\
   o_2 &= 2l - e_1 - \dots - e_6 = f - n_2 .
\end{aligned}
\label{eq:no}
\end{equation}
Under $\tP$ the reducible $I_2$ fibers $f_1$ and $f_2$ must be
exchanged. Thus, $\t_B$ must somehow exchange $(n_1,o_1)$ and
$(n_2,o_2)$. Specifically, one finds  
\begin{equation}
\begin{aligned}
   \t_B^*(n_1) &= o_2 , \\
   \t_B^*(o_1) &= n_2 . \\
\end{aligned}
\label{eq:tBno}
\end{equation}
The full action of $\t_B^*$ on the cohomology is given in
Table~\ref{tab:tB}. 

This completes the description of the family of rational elliptic
surfaces and the involutions $\t_B$ which we will use to construct $(X,\t_X)$.

\subsection{Construction of $(X,\t_X)$}
\label{sec:X}

Given our specific family of rational elliptic surfaces $B$, we can
now describe the construction of a suitable family of elliptically fibered
$(X,\t_X)$. The fact that rational elliptic surfaces are themselves
elliptically fibered, allows a particularly simple construction of $X$
as the fiber product, 
\begin{equation}
   X = B \times_{\cp{1}} B' ,
\label{eq:Xfiber}
\end{equation}
of a pair of rational elliptic surfaces $B$ and $B'$. That is to say,
$X$ fits into a commutative diagram of projections
\begin{equation}
  \xymatrix{
       & X \ar[dl]_-{\p'} \ar[dr]^-{\p} & \\
     B \ar[dr]_-{\b} & & B' \ar[dl]^-{\b'} \\
       & \cp{1} &
  } 
\label{eq:Xfibration}
\end{equation}
The space $X$ is formed by taking the union, as $p$ varies in
${\mathbb P}^{1}$, of the product of the fibers
$\b^{-1}(p)\times {\b'}^{-1}(p)$ above $p$. This
is shown schematically in Figure~\ref{fig:X}. 
\FIGURE{%
   \parbox{\textwidth}{%
   \begin{center}
      \epsfig{file=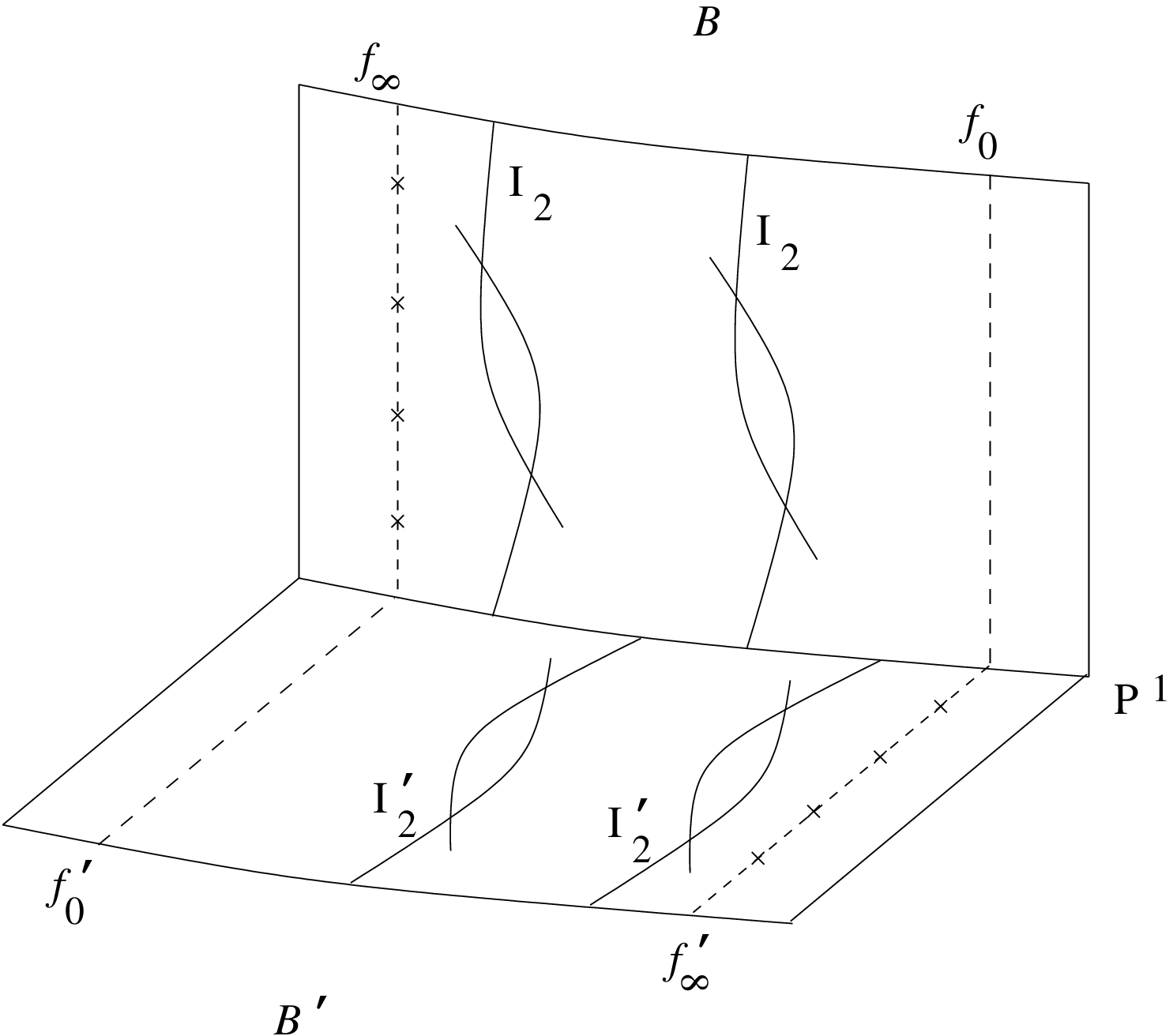,height=3.0in}
   \end{center}
   \caption{The structure of $X$.}
   \label{fig:X}
}}

For generic choice of $B$ and $B'$, $X$ will be smooth. It is an
elliptic fibration in two ways: either via $\p$ or $\p'$. Since most
of our construction will center on the elliptic fibers, we will make
the somewhat unconventional choice that the primed object $B'$ is the
base of the fibration, simply to avoid cumbersome notation. Much of
the structure of the fibration is inherited from the structure of the
$\b$ fibration of $B$. For instance, the discriminant of $\p$ is the
pull-back of the discriminant of $\b$ and so is a section of
$\cO(12f')=K_{B'}^{-12}$. Hence, $c_1(X)=0$ and $X$ is
Calabi--Yau. Similarly, the zero section $\s:B'\to X$ of $\p$ is
inherited from the zero section $e:\cp{1}\to B$, and is given by
$\s=e\times_{\cp{1}}B'$. 

Let us now assume that $B$ and $B'$ are special in the sense of
section~\ref{sec:specialB}. We then have involutions $\a_B$, $\t_B$
and $\a_{B'}$, $\t_{B'}$ acting on $B$ and $B'$ from which we will
construct $\t_X$. There is still some freedom in how we choose to
identify the $\cp{1}$ bases of $B$ and $B'$. However, since all
involutions of $X$ must induce the same involution $\tP$, there are
only two possibilities, depending on the identification of fixed
points. Either we identify $0\in\cp{1}$ with $0'\in{\cp{1}}'$ and
$\infty\in\cp{1}$ with $\infty'\in{\cp{1}}'$, or $0$ with $\infty'$
and $\infty$ with $0'$. Suppose we make the first
identification. Recall that both $\a_B$ and $\t_B$ leave four points
on $f_\infty$ fixed. Since in the first case, both $f_\infty$ and
$f'_\infty$ lie above the same point in $\cp{1}$, it is clear that no
combinations of involutions in $B$ and $B'$ can be freely
acting. However, with the second identification it is easy to see that
the involution 
\begin{equation}
   \t_X = \t_B \times_{\cp{1}} \t_{B'}
\label{eq:tX}
\end{equation}
acts freely on $X$. This is because the four fixed points of
$\t_B$ and $\t_{B'}$ live in fibers above different points in
$\cp{1}$, as shown in Figure~\ref{fig:X}. Note that this would never
be the case if $\t_X$ was built from either $\a_B$ or $\a_{B'}$. 

As for the involutions of the rational elliptic surfaces, the
involution $\t_X$ can also be built from an involution $\a_X$
preserving the zero section $\s$ and a translation $t_{\z_X}$ by a
second section $\z_X$. (This was described in~\cite{z2}.) Both are
built out of the corresponding structures on $B$, namely 
\begin{equation}
   \a_X = \a_B \times_{\cp{1}} \t_{B'}, 
\label{eq:aX}
\end{equation}
and 
\begin{equation}
   \label{eq:zX}
   \z_X = \z \times_{\cp{1}} B' . 
\end{equation}

In what follows, it will be useful to identify the classes of divisors
$H^2(X,\ZZ)$ on $X$. The fiber product structure means that 
\begin{equation}
   H^2(X,\ZZ) = \frac{H^2(B,\ZZ)\times H^2(B',\ZZ)}{H^2(\cp{1},\ZZ)}
\label{eq:H2X}
\end{equation}
That is, all divisor classes are either pull-backs of classes from $B$
or of classes from $B'$, modulo the one relation on the fiber classes
that $\p^*(f')={\p'}^*(f)$. 

Finally, we will also need the expression for $c_2(X)$ in order to
solve the condition~\ref{C2}. In general, this is given by a curve in
$X$. Recalling that $c_2(B)=c_2(B')=12$ and the fibered structure of
$X$, it is easy to show that $c_2(X)$ is given solely by some number
of $\p$ and $\p'$ fibers. That is 
\begin{equation}
   c_2(X) = 12\left(f\times\pt + \pt\times f'\right) ,
\label{eq:c2X}
\end{equation}
where $\pt$ is the class of a point in the relevant manifold ($B$ or
$B'$ or later $X$). 

In summary, we have described the construction of 
 a large class of threefolds $X$ with
freely acting $\t_X$. The quotient $Z=X/\t_X$ will thus be smooth, with
non-trivial $\p_1(Z)$. We note that it is not difficult to show that
$Z$ is also a Calabi--Yau manifold, as required.


\section{A Family of $\t_X$-Invariant Bundles $V$}
\label{sec:V}

In this section, we will describe the construction 
a large family of stable bundles 
$V$ on $X$ which are invariant under the involution $\t_X$. 
As in previous papers~\cite{dlow1,dlow2,moduli,z2}, we will use 
the spectral construction. One new feature is that we are forced to
work with a reducible spectral cover. Rather than describing the
complicated behavior of the spectral sheaf at the singularities of the
spectral cover, we chose to realize the resulting vector bundle as an
extension of two vector bundles $V_{2}$ and $V_{3}$ each coming from more
manageable spectral data. This approach is a variation of an idea of
Richard Thomas \cite{richardt}.

The key ingredient in the spectral construction is the
fact that $X$ is elliptically fibered. This allows $V$ to be
constructed via the ``Fourier--Mukai transform'' which, in physics
terms, is the action on the bundle of T-duality along the elliptic
fibers. Recall that, with respect to the complex structure, the T-dual
Calabi--Yau manifold is isomorphic to $X$. Formally, the
Fourier--Mukai transform $\FM_X$ is then an ``autoequivalence'' of the
``derived category'' $D^b(X)$ of sheaves on $X$, 
\begin{equation}
   \FM_X : D^b(X) \to D^b(X) .
\label{eq:FMX}
\end{equation}
Physically, we can think of a sheaf as describing a D-brane. Thus, as
expected, $\FM_X$ maps one configuration of D-branes to another. It is
really a little subtler. The objects in $D^b(X)$ are actually not
single sheaves but complexes of sheaves. Although very important for
the details of the construction (see~\cite{math}), this subtlety will
not generally concern us here. 

The usefulness of the Fourier--Mukai transform is that it allows 
one to describe $V$ in terms of its simpler T-dual data. In particular 
consider a line bundle $\cN_\S$ over a smooth surface 
$i_\S:\S\hookrightarrow X$ in $X$ which is a finite $r$-fold cover of the 
base $B'$. The transform of the corresponding sheaf $i_{\S*}\cN_\S$ on $X$, 
\begin{equation}
   V = \FM_X(i_{\S*}\cN_\S) ,
\label{eq:spectral}
\end{equation}
is then precisely the object we want: a stable vector bundle 
over $X$ of rank $r$. (We should note that the stability also depends
on a choice of a suitable K\"ahler form on $X$.) The surface $\S$ is
the spectral cover, $\cN_\S$ the spectral datum and  the
correspondence between $(\S,\cN_\S)$ and $V$ is commonly known as the
spectral construction.  

In order to find $\t_X$-invariant bundles, we would like to translate the 
invariance condition into a condition on $(\S,\cN_\S)$. Since $\FM_X$ 
is invertible, we can construct the induced action of $\t_X$ on 
the spectral sheaf $i_{\S*}\cN_\S$ 
\begin{equation}
   \T_X = \FM_X^{-1} \circ \t_X^* \circ \FM_X .
\label{eq:TX}
\end{equation}
The search for invariant bundles $V$ is then reduced to finding 
$(\S,\cN)$ such that 
\begin{equation}
   \T_X(i_{\S*}\cN_\S) \cong i_{\S*}\cN_\S .
\label{eq:TXcond}
\end{equation}

In general, the action of $\T_X$ is extremely hard to calculate. One 
particular problem is that the space $\Pic(\S)$ of line bundles
$\cN_\S$ on $\S$ is generically not simply characterized by pullbacks
of bundles from $X$. Instead, as we saw in~\cite{z2}, new divisor
classes on $\S$ appear. Calculating $\T_X$ for the corresponding
bundles is difficult. However, here, the fiber-product structure of
$X$ will come to our aid. Just as many of the properties of $X$ were
inherited from the vertical surface $B$, in the cases of interest, we
will be able to build the Fourier--Mukai transform $\FM_X$ from the
horizontal pullback by ${\p'}^*$ of the corresponding transform
$\FM_B$ on $B$. 

Our first step in this section is, thus, to give some results on the
action of $\FM_B$ and the corresponding $\T_B$ on the special rational
elliptic surfaces $B$. We then use these results, together with the
technique of modifying bundles by ``Hecke transforms,'' to construct a
large class of $\t_X$-invariant bundles.

\subsection{The Fourier--Mukai Transform on $B$ and a No-Go Theorem}
\label{sec:global}

Since $B$ is elliptically fibered, there is also a Fourier--Mukai
action $\FM_B$ on sheaves on $B$. Similarly, given the involution
$\t_B$ on $B$, there is an induced action
$\T_B=\FM_B^{-1}\circ\t_B^*\circ\FM_B$ on sheaves on $B$. Let us now
simply state some results for $\FM_B$ and $\T_B$. Details can
be found in~\cite{math}. 

Our main result is the following. Let $L$ be a line bundle on $B$. In
general $\FM_B(L)$ will be some complex of sheaves. Nonetheless, one
can show, as long as $c_1(L)\cdot o_i=0$, that $\T_B(L)$ is actually
still a line bundle. Explicitly, in terms of the divisor classes
defined in section~\ref{sec:specialB}, one shows (see
\cite[Part~I,Theorem~7.1]{math}):  
\begin{equation}
   \T_B(L) = \t_B^*(L) \otimes 
      \cO\bigl([c_1(L)\cdot(e-f)]f + [c_1(L)\cdot f](e-\z-f)\bigr)
      \otimes \cO(e-\z-f) . 
\label{eq:TBL}
\end{equation}
Thus we see that $\T_B$ induces a complicated affine action on
the space of line bundles $\Pic(B)$. The first two terms give the
underlying linear part of the transformation, while the last term gives
the constant shift. 

The analogue in $B$ of the spectral cover $\S$ is a smooth curve
$i_C:C\hookrightarrow B$ which is a finite cover of the base
$\cp{1}$. The analog of $\cN_\S$ is then a line bundle $\cN$ on
$C$. Consequently, we would also like to know the action of $\T_B$ on
spectral sheaves $i_{C*}\cN$. Let us assume that the bundle $\cN$ is
the pullback $\cN=i_C^*(L)$ of a global bundle $L$ in $B$ (note that,
in general, this is not always the case). One can then show that
\begin{equation}
   \T_B(i_{C*}i_C^*(L)) = i_{D*}i_D^*(\T_B(L)) , 
\label{eq:TBCL}
\end{equation}
where
\begin{equation}
   D = \a_B(C)
\label{eq:Ddef}
\end{equation}
is the image of $C$ under the involution $\a_B$. This matches the
result of~\cite{z2}, where it was shown that the spectral cover
transforms under the involution preserving the zero section, $\a_B$,
rather than the full involution, $\t_B$, of the manifold. 

Finally, if we restrict $C$ to be in the divisor class $re+kf$
for some integers $r$ and $f$, one specifically finds,
using~\eqref{eq:TBL}, that   
\begin{equation}
   \T_B(i_{C*}i_C^*(L)) = i_{D*}i_D^*\left( 
       \a_B^*(L) \otimes \cO(e-\z-f) \right) .
\label{eq:TBspecial}
\end{equation}

We can now use these results for $\T_B$ to show a useful no-go theorem
for constructing suitable bundles on $X$. The most obvious
simplification in the spectral construction is to ignore any new
classes on the spectral cover $\S$, and to assume that the line bundle
$\cN_\S$ is the pullback of a global line bundle $\cL$ on $X$
\begin{equation}
   \cN_\S = i_\S^*(\cL) . 
\label{eq:Nglobal}
\end{equation}
Recall from equation~\eqref{eq:H2X} that all the divisor classes on
$X$ came as either pullbacks of classes on $B$ or pullbacks of classes
on $B'$. Thus, in general, $\cL$ can be written as 
\begin{equation}
   \cL = {\p'}^*L \otimes \p^*L' ,
\label{eq:cL}
\end{equation}
where $L$ and $L'$ are global line bundles on $B$ and $B'$
respectively. The action of $\FM_X$ then splits into a Fourier-Mukai 
action on $B$
and the trivial action on $B'$. Specifically
\begin{equation}
   \FM_X(\cL) = {\p'}^*\FM_B(L) \otimes \p^*L' . 
\label{eq:FMglobal}
\end{equation}
Similarly, given the form~\eqref{eq:tX} of $\t_X$, the action of
$\T_X$ is given by  
\begin{equation}
   \T_X(\cL) = {\p'}^*\T_B(L) \otimes \p^*\t_{B'}^*L'
\label{eq:Tglobal}
\end{equation}
One notes that the action of $\T_X$ is simple on $L'$ but more
complicated on $L$. This reflects the fact that the Fourier--Mukai
transformation is the action of T-duality on the $\p$ fibers. 

From these relations it is straight-forward to deduce the action
of $\T_X$ on the spectral data $(C,i_C^*(\cL))$. The bundle is
invariant under the involution provided that 
\begin{equation}
\begin{aligned}
   \a_X(\S) &= \S , \\
   \t_{B'}^* L' &\cong L' , \\
   \T_B(L) &\cong L
\end{aligned}
\label{eq:invglobal}
\end{equation}
Finding solutions of the first two conditions it relatively easy. We
can then use the general result~\eqref{eq:TBL} to try and solve the
$L$ condition. Rewriting the action of $\T_B$ in terms of cohomology,
we see that~(Prop.~2.11 in~\cite{math}).  
\begin{equation}
   c_1(\T_B(L)) = \t_B^*(c_1(L)) 
      + \left[c_1(L)\cdot(e-f)\right] f 
      + \left[c_1(L)\cdot f+1\right] (e-\z+f) .
\label{eq:TBLcohom}
\end{equation}
Using the action of $\t_B^*$ given in Table~\ref{tab:tB}, it is easy
to see that $c_1(\T_B(L))=c_1(L)$ if and only if $c_1(L)$ is in the
affine subspace of $H^2(B,\bbq)$
\begin{equation}
   -\frac{1}{2}e_1 
     + \Span(f,e_9,e_4-e_5,e_4-e_6,l-e_7-2e_8,3l-2(e_4+e_5+e_6)-3e_7) .
\label{eq:span}
\end{equation}
However, $c_1(L)$ must be in $H^2(B,\bbz)$ and there are no integral
vectors in this subspace. Thus, we see that there are no solutions to
the $L$ condition. 

We have derived a no-go theorem: $V$ can never be $\t_X$-invariant if
$\cN_\S$ is the pullback of a global line bundle on $X$. Instead we
are forced to consider cases where $\cN_\S$ comes at least partly from
additional classes on $\S$.

\subsection{Construction of $V_r$}
\label{sec:Vr}

Given the general results of the last section, we can now turn to the
specific construction of a suitable family of rank $r$ bundles $V_r$. We
start by defining an appropriate spectral cover $\S$. Again, we can
use the projection $\p'$ to describe it as a pullback of a simpler
object in $B$. Let $C$ be a smooth irreducible curve in $B$ suitable
for a spectral construction in $B$. Specifically, let $C$ be
irreducible and in the divisor class  
\begin{equation}
   [C] = re + kf ,
\label{eq:C}
\end{equation}
for some integer $k$, so that it is an $r$-fold cover of the base
$\cp{1}$. We then take $\S$ to be the pullback 
\begin{equation}
   \S = C \times_{\cp{1}} B' . 
\label{eq:Sigma}
\end{equation}
Given the properties of $C$, the spectral cover $\S$ will be a smooth, 
irreducible $r$-fold cover of $B'$. By construction, $\S\to C$ is an
elliptic fibration over $C$. In particular, it includes some number of
reducible fibers. Let $f_1'=n'_1\cup o'_1$ and $f_2'=n'_2\cup o'_2$ be
the reducible fibers of $B'$ (as in equation~\eqref{eq:I2}). Let $F_1$
and $F_2$ be the fibers of $B$ over the corresponding points in
$\cp{1}$. This is shown schematically in Figure~\ref{fig:S}. 
\FIGURE{%
   \parbox{\textwidth}{%
   \begin{center}
      \epsfig{file=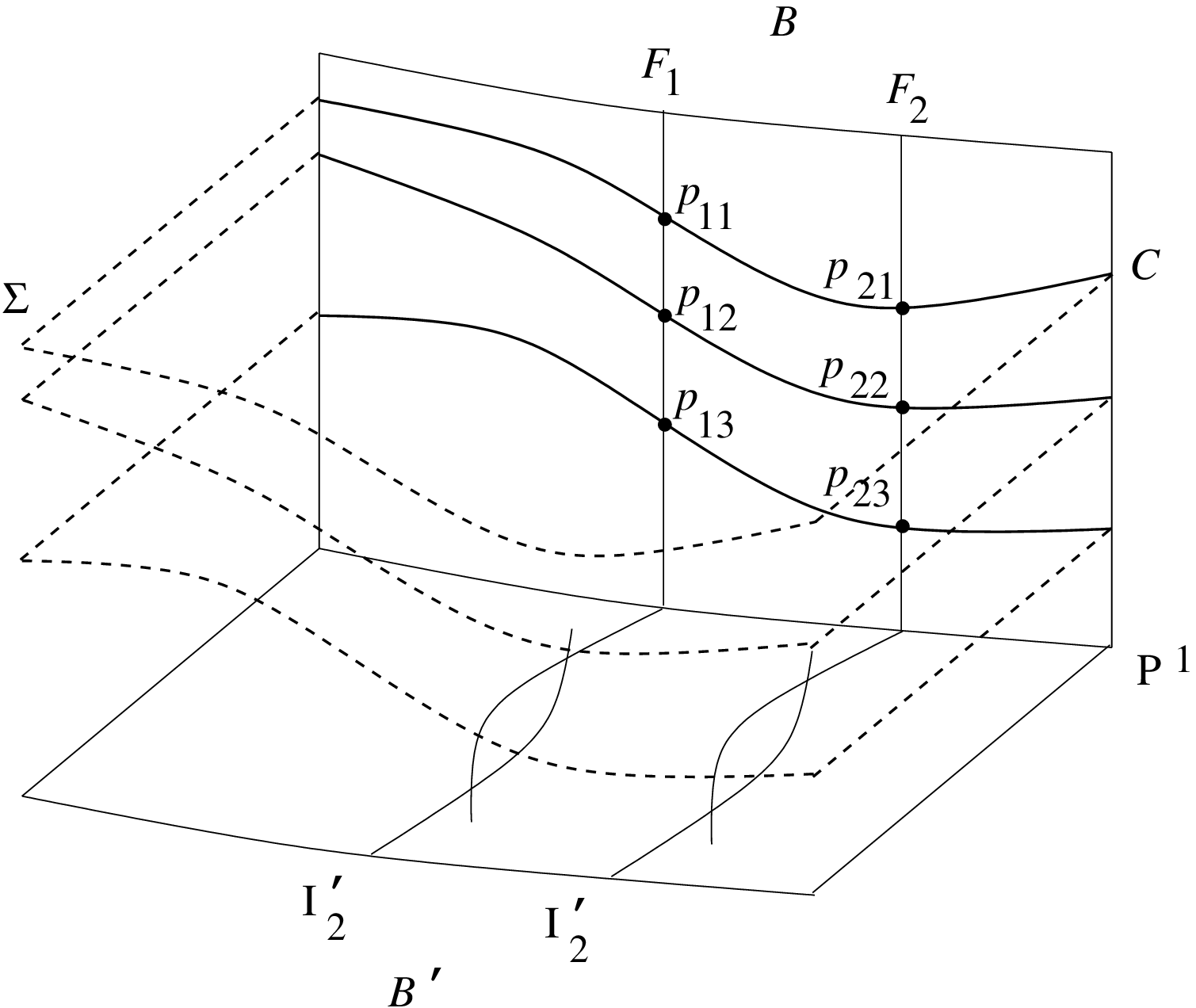,height=3.0in}
   \end{center}
   \caption{The structure of the spectral cover $\S$.}
   \label{fig:S}
}}
Given the way we glued the $\cp{1}$ bases of $B$ and $B'$, the fibers
$F_1$ and $F_2$ are smooth. The fibers $F_1$ and $F_2$ will each
intersect the curve $C$ in $r$ points in $B$. Let us label
these by an index $\k=1,\dots,r$, so  
\begin{equation}
   C \cap F_j = \left\{ p_{j\k} \right\}_{\k=1}^r
\label{eq:p}
\end{equation}
for $j=1,2$. Above each of these points, the fiber of $\S$ will
split. Thus $\S$ is an elliptic surface with $2r$ fibers of type $I_2$
given by $(n'_j\cup o'_j)\times \{p_{j\k}\}$ for $j=1,2$ and
$\k=1,\dots,r$. 

Next, we turn to the line bundle $\cN_\S$. Generically, there are three
types of divisor on $i_\S:\S\hookrightarrow X$. First, there are
pullbacks under $i_\S^*$ of global divisors on $X$. Then, we
have pullbacks of divisors (points) on $C$ and, finally, the $2r$ new
divisors coming from the reducible fibers. Thus, we can take $\cN_\S$
of the form
\begin{equation}
   \label{eq:cNS}
   \cN_\S = {\p'_{|\S}}^* \cN 
      \otimes \cO_\S\left( -\sideset{}{_{j\k}}\sum 
           \{p_{j\k}\}\times(a_{j\k}n'_j+b_{j\k}o'_j) \right)
      \otimes i_\S^*\p^*L ,
\end{equation}
where $\cN$ is a line bundle of degree $d$ on $C$, $a_{j\k}$ and
$b_{j\k}$ are integers and $L$ is a line bundle on $B'$. The first
term is precisely the pullback of a bundle on $C$. The second is
a bundle corresponding to some combination of the new divisors from
the reducible fibers. The last term is the pullback of a global bundle
on $X$. We note that there is some redundancy
in the choices of $a_{j\k}$ and $b_{j\k}$. Since $n'_j+o'_j=f'_j$ is a
pullback from $\cp{1}$, it can be absorbed in $L$. Thus, we are free to 
take 
\begin{equation}
   \text{either $a_{j\k}\geq0$, $b_{j\k}=0$ or $a_{j\k}=0$,
      $b_{j\k}\geq0$ for all $j$ and $\k$} . 
\label{eq:ab}
\end{equation}

Finally, we should stress that both the pullback of $\cN$ and the
bundles from the extra reducible fibers represent contributions to
$\cN_\S$ which are not pullbacks of global line bundles on
$X$. Thus, we can hope to avoid the no-go theorem of
section~\ref{sec:global}. In fact, generalizing to include $\cN$ would
be sufficient to find invariant bundles. However, as we will see, in
order to satisfy conditions~\ref{S}--\ref{C3}, one needs the
additional freedom of the reducible fibers.  

Having defined $(\S,\cN_\S)$, we now need to understand the action of
$\FM_X$ and $\T_X$. This could be addressed directly on $X$. However,
in fact, the action can be decomposed in the following, relatively
simple way. We first note that, as in equation~\eqref{eq:FMglobal}
above, we can factor off the contribution of the global line bundle
$\p^*L$ under the action of $\FM_X$. We have 
\begin{equation}
   V_r = \FM_X(i_{\S*}\cN_\S) = \tW \otimes \p^*L ,
\label{eq:tW}
\end{equation}
where $\tW$ is the bundle constructed from $\S$ and the spectral
datum $\tN_\S$ given by the first two terms
in~\eqref{eq:cNS}. If, somehow, we could also remove the contribution
from the reducible fibers, we would then be left with a bundle which,
given the structure of $\S$, is just a pullback of a bundle $W$ on $B$
(similar to the case of equation~\eqref{eq:FMglobal}),
\begin{equation}
   \FM_X(i_{\S*}{\p'_{|\S}}^*\cN) = {\p'}^*W ,
\label{eq:reduce}
\end{equation}
where,
\begin{equation}
   W = \FM_B(i_{C*}\cN) .
\label{eq:W}
\end{equation}
This can then be calculated given our results on $\FM_B$ in
section~\ref{sec:global}.

It turns out that there is a precise way to go from ${\p'}^*W$ to
$\tW$. It is the action of a series of Hecke transforms. Details
can be found in~\cite{math}. Here we will simply note that, given a
vector bundle $E$ on $X$, a divisor $i_D:D\hookrightarrow X$ and a
short exact sequence $(\xi) : 
0\to F\to E_{|D}\to G\to0$ of vector bundles on
$D$, the associated Hecke transform generates a new vector bundle
$\Hecke_{(\xi)}(E)$  on
$X$. This new bundle has two characteristic properties: the Chern
character of $\Hecke_{(\xi)}(E)$ is equal to $\ch(E) - \ch(i_{D*}G)$,
and  $\Hecke_{(\xi)}(E)$ is isomorphic to $E$ on the complement of $D$.
\begin{equation}
   \tW = \Hecke_{a_{j\k},b_{j\k}}({\p'}^*W) ,
\label{eq:Hecke}
\end{equation}
where $\Hecke_{a_{j\k},b_{j\k}}$ represents $a_{j\k}$ successive Hecke
transforms on the the divisor $D=F_j\times n'_j$ together with
$b_{j\k}$ successive Hecke transforms on $D=F_j\times o'_j$. 

In summary, we have built $V_r$ as 
\begin{equation}
   V_r = \Hecke_{a_{j\k},b_{j\k}}({\p'}^*(\FM_B(i_{C*}\cN)))
      \otimes \p^*L .
\label{eq:Vr}
\end{equation}
In the remainder of this section we want 
to show that there are suitable $a_{j\k}$, $b_{j\k}$,
$L$ and $\cN$ such that $V_r$ is invariant under $\t_X$. 

Acting with $\t_X$ on $V_r$, it is clear from equation~\eqref{eq:tW}
and the form~\eqref{eq:tX} of $\t_X$ that 
\begin{equation}
   \label{eq:tXVr}
   \t_X^*(V_r) = \t_X^*(\tW) \otimes \p^*\t^*_{B'}L . 
\end{equation}
Thus, as in equation~\eqref{eq:invglobal}, invariance of $V_r$
requires 
\begin{equation}
   \t_{B'}L \cong L ,
\label{eq:Lcond}
\end{equation}
and 
\begin{equation}
   \t_X^* \tW \cong \tW .
\label{eq:Wtcond}
\end{equation}
Recall that the Hecke transforms were on the divisors $F_j\times n'_j$
and $F_j\times o'_j$. From~\eqref{eq:tBno}, we see that $\t_{B'}$
exchanges $n'_1$ with $o'_2$ and $n'_2$ with $o'_1$. Hence, for
invariance under $\t_X$ we must perform the same number of Hecke
transforms on each set of divisors paired under $\t_{B'}$. This
implies that $a_{1\k}=b_{2\k}$ and
$a_{2\k}=b_{1\k}$. Given~\eqref{eq:ab}, we  take 
\begin{equation}
   a_{1\k} = b_{2\k} \cong a_{\k} \geq 0 , \qquad
   a_{2\k} = b_{1\k} = 0 .
\label{eq:abcond}
\end{equation}
After undoing all the Hecke transforms, the $\t_X$-invariance of
$\tW$ reduces, from equation~\eqref{eq:reduce}, to the
$\t_X$-invariance of ${\p'}^*W$. Since this is just the pull back of a
bundle from $B$, given the expression~\eqref{eq:W}, we finally have the
condition
\begin{equation}
   \T_B(i_{C*}\cN) \cong i_{C*}\cN .
\label{eq:Wcond}
\end{equation}
Since $[C]=re+kf$, using the result~\eqref{eq:TBspecial} for global
line bundles on $B$, one can show that equation~\eqref{eq:Wcond}
implies that 
\begin{align}
   C &= \a_B(C) , 
      \label{eq:Ccond} \\
   \cN &\cong \a_{B|C}^*(\cN) \otimes \cO_C(e-\z+f) .
      \label{eq:Ncond}
\end{align}
where $\a_{B|C}$ is the restriction of the involution $\a_B$ to $C$. 

The requirement that $\t_X^*V_r\cong V_r$, has been reduced to the four
conditions~\eqref{eq:Lcond}, \eqref{eq:abcond}, \eqref{eq:Ccond} and
\eqref{eq:Ncond}. From Table~\ref{tab:tB}, we see that there is a
six-dimensional lattice of $\t_{B'}$-invariant classes on $B'$, so
there are many possibilities for $L$. The conditions on $a_{j\k}$ and
$b_{j\k}$ simply reduce to the choice of positive integers
$a_{\k}$. Since the set of divisors $C$ in the class $re+kf$ forms a
projective space, there must be a solution to the condition on $C$. In
general, $C$ will be smooth provided
\begin{equation}
   k \geq r > 1 .
\label{eq:rkcond}
\end{equation}
However, it is not obvious that the general $C$ satisfying
\eqref{eq:Ccond} will be smooth as well. In fact, for the values of
$k$ which we need, this turns out to be true for $r = 3$ and false for
$r = 2$. The relevant curves $C_{2}$ are actually reducible and always
have a vertical component. This requires a separate analysis of
the invariance properties of $W_{2}$ which is carried out in \cite{math}.
In \cite{math} it is also shown that for all bundles
$\cN$ on $C$ there is some positive-dimensional torus of solutions to
the condition~\eqref{eq:Ncond}. 

We conclude, therefore, that using the construction~\eqref{eq:Vr}, we
can build a large family of $\t_X$-invariant rank-$r$ bundles $V_r$ on
$X$.

\subsection{Extensions and Stability}
\label{sec:ext}

As we shall see in section~\ref{sec:num}, it turns out that even with
the additional freedom coming from the reducible classes on $\S$, the
bundles $V_r$ that we have just constructed are not quite general
enough to satisfy all the conditions \ref{S}--\ref{C3}. We will
actually construct our rank-five bundle $V$ as the extension 
\begin{equation}
   0 \to V_2 \to V \to V_3 \to 0 ,
\label{eq:extension}
\end{equation}
where $V_r$, with $r=2,3$, are $\t_X$-invariant rank-$r$ bundles
constructed via the procedure given above, and the extension class
itself is also $\tau_{X}$-invariant. 

By construction, $V$ will be $\t_X$-invariant. However, we also
require~\ref{S} that $V$ be stable given a suitable K\"ahler form
$H$ on $X$. This puts some constraint on the extension and also on the
bundle $V_2$. Again, we will simply quote the result. Stability of $V$
is equivalent, first, to the fact that $V$ is not split, that is, it is
not a direct sum,  
\begin{equation}
   V \neq V_2 \oplus V_3
\label{eq:Vcond1}
\end{equation}
and, second, to a condition on the slopes 
\begin{equation}
   \m(V_2) < \m(V) 
\label{eq:Vcond2}
\end{equation}
where, for an arbitrary bundle $E$, $\m(E)=\left(\int_X c_1(E)\wedge
H\wedge H\right)/\rk(E)$. (Note that condition~\ref{C1} implies
that $\m(V)=0$.)


\section{Numerical Conditions}
\label{sec:num}

In the previous section, we built a class of bundles $V$ satisfying
the $\t_X$-invariance condition~\ref{I}. We now need to find the
requirements on $V$ for simultaneously satisfying all the
conditions~\ref{S}--\ref{C3}. In particular, we will reduce them to a
set of numerical constraints on the parameters defining $V$. 

Recall that the invariance conditions~\eqref{eq:Ccond}
and~\eqref{eq:Ncond} were geometrical in nature, fixing a particular
$C$ and $\cN$. Let us assume these conditions are satisfied. Recalling
that $V$ is built from two bundles $V_r$, for $r=2,3$, the bundle is
then determined by the following parameters, again indexed by $r$,
\begin{itemize}
\item the integers $k_r$ giving the classes of the curves $C_r$ as
   in~\eqref{eq:C},
\item the integers $d_r=\deg(\cN_r)$ giving the degrees of the bundles
   $\cN_r$, 
\item the integers $a_{r\k}$ with $\k=1,\dots,r$ determining the
   number of Hecke transforms~\eqref{eq:abcond}, 
\item and the line bundles $L_r$ on $B'$.
\end{itemize}

The $\t_X$-invariance condition only constrains $L_r$. We have
\begin{quote}
\begin{enumerate}
\renewcommand{\labelenumi}{\textbf{(\#I)}}
\renewcommand{\theenumi}{\textbf{(\#I)}}
\item \label{nI} \quad
   $\t_B^*L_r = L_r$, for $r=2,3$.
\end{enumerate}
\end{quote}
Recall that there was also a condition~\eqref{eq:rkcond} in the
construction of $V_r$ in order for a generic 
$C_r$ to be smooth. Here, it implies
\begin{equation}
   k_2 \geq 2 \text{ and } k_3 \geq 3. 
\label{eq:kcond}
\end{equation}

Let us now turn to the condition of stability~\ref{S}. We saw that
this implied that $V$ did not split~\eqref{eq:Vcond1} and a condition
on the slopes~\eqref{eq:Vcond2}. It can be shown~\cite{math} that
these conditions amount to
\begin{quote}
\begin{enumerate}
\renewcommand{\labelenumi}{\textbf{(\#Se)}}
\renewcommand{\theenumi}{\textbf{(\#Se)}}
\item \label{nSe} \quad
   $L_2\cdot f' > L_3\cdot f'$,
\renewcommand{\labelenumi}{\textbf{(\#Ss)}}
\renewcommand{\theenumi}{\textbf{(\#Ss)}}
\item \label{nSs} \quad
   $(2L_2+(d_2-2k_2+1)f' - S_2^1(n_1'+o_2')) \cdot h' < 0$ 
   for some ample class $h' \in H^2(\bbz,B')$,
\end{enumerate}
\end{quote}
respectively. Here we have introduced the notation $S^p_r$ for sums of
$p$-th powers of $a_{r\k}$ 
\begin{equation}
   S^p_r = \sum_{\k=1}^r \left(a_{r\k}\right)^p. 
\label{eq:Sp}
\end{equation}

What remains are the conditions~\ref{C1}--\ref{C3} on the Chern
classes. Using the explicit construction~\eqref{eq:Vr}, one can derive
the following expression for $\ch(V_r)$ for $V_2$ and $V_3$, 
\begin{equation}
\begin{split}
   \ch(V_r) &= r 
      + \p^*\left(
          rL_r + \left(d_r - rk_r + \binom{r}{2}\right)f' 
      - S_r^1 (n'_1 + o'_2) \right) 
      \\ & \quad
      + \left[ \frac{r}{2}L_r^2 
          + \left(d_r - rk_r + \binom{r}{2}\right)L_r\cdot f' 
          - S_r^1\left(L_r\cdot n'_1 + L_r\cdot o'_2\right)
          - 2S_r^2 \right]
          (f \times \pt)
      \\ & \quad
      - k_r (\pt \times f') 
      - k_r (L_r\cdot f') \pt .
\label{eq:chVr}
\end{split}
\end{equation}
Note that $c_1(V_r)$ is a pullback from $B'$, while the only terms in
$\ch_2(V_r)$ are proportional to $f \times \pt$ and $\pt \times
f'$. More significantly, one notes that requiring $c_1(V_r)=0$ implies
that $L_r\cdot f'=0$ in $B'$. From the last term in~\eqref{eq:chVr}
this, in turn, implies that $c_3(V_r)$ vanishes. Clearly, $V_r$ by
itself cannot, therefore, satisfy both conditions~\ref{C1}
and~\ref{C3}. This is the reason we were forced to consider the
generalization of $V$ constructed as an extension. 

Given the form of $V$, we have $\ch(V)=\ch(V_2)+\ch(V_3)$. Combining
this with~\eqref{eq:chVr} and~\eqref{eq:c2X}, the
conditions~\ref{C1}--\ref{C3} imply the following numerical
constraints 
\begin{quote}
\begin{enumerate}
\renewcommand{\labelenumi}{\textbf{(\#C1)}}
\renewcommand{\theenumi}{\textbf{(\#C1)}}
\item \label{nC1} \quad
   $2L_2+3L_3 = (S_2^1+S_3^1)(n_1'+o_1') - (d_2+d_3-2k_2-3k_3+4)f'$,
\renewcommand{\labelenumi}{\textbf{(\#C2$f$)}}
\renewcommand{\theenumi}{\textbf{(\#C2$f$)}}
\item \label{nC2f} \quad
   $k_2 + k_3 \leq 12$
\renewcommand{\labelenumi}{\textbf{(\#C2$f'$)}}
\renewcommand{\theenumi}{\textbf{(\#C2$f'$)}}
\item \label{nC2fp} \quad
   $L_2^2 + \frac{3}{2}L_3^2 + (d_2-2k_2+1)(L_2\cdot f') 
      + (d_3-3k_3+3)(L_3\cdot f') - (S_2^1L_2+S_3^1L_3)(n_1'+o_2')
      - 2(S_2^2+S_3^2) \geq -12$,
\renewcommand{\labelenumi}{\textbf{(\#C3)}}
\renewcommand{\theenumi}{\textbf{(\#C3)}}
\item \label{nC3} \quad
   $k_2(L_2\cdot f') + k_3(L_3\cdot f') = -6$.
\end{enumerate}
\end{quote}
Note that the $c_2(V)$ condition splits into two pieces, one
from the component proportional to $f\times\pt$ and one from that
proportional to $\pt\times f'$.


\section{A Class of Solutions}
\label{sec:examples}

What remains is to find a simultaneous solution of the
equations~\ref{nI}, \ref{nSe} and~\ref{nSs},
and~\ref{nC1}--\ref{nC3} together with the
inequality~\eqref{eq:kcond}. It is a straightforward, if tedious,
procedure to calculate a fairly general solution~\cite{math}. Let us
summarize the result, noting only that the main constraint on finding
solutions is the tension between the stability condition~\ref{nSe} and
the $c_2(V)$ conditions~\ref{nC2f} and~\ref{nC2fp}{}. 

First, solving conditions~\ref{nC1}, \ref{nC3} and~\ref{nI},
constrains the line bundles $L_r$ to have the following form 
\begin{equation}
\begin{aligned}
   L_2 &= \frac{9}{k}\left(e' + \z'\right) 
      + \frac{1}{2}\left(x - d_2 + 2k_2 - 1\right)f'
      + \frac{1}{2}\left(u + \frac{9}{k} + S_2^1\right)
          \left(n_1'+o_2'\right)
      + 3M  \\
   L_3 &= -\frac{6}{k}\left(e' + \z'\right) 
      + \frac{1}{3}\left(-x - d_3 + + 3k_3 - 3\right)f'
      + \frac{1}{3}\left(-u - \frac{9}{k} + S_3^1\right)
          \left(n_1'+o_2'\right)
      - 2M  \\
\end{aligned}
\label{eq:Lsol}
\end{equation}
where $k=2k_3-3k_2$ and $x$ and $u$ are as yet undetermined. From
Table~\ref{tab:tB} and equation~\eqref{eq:tBno}, we note that
$e'+\z'$, $f'$ and $n_1'+o_2'$ are $\t_B$-invariant classes. The
parameter $M$ represents an  arbitrary $\t_B$ invariant class
orthogonal to $e'+\z'$, $f'$ and $n_1'+o_2'$. It is then clear that
the $L_r$ satisfy~\ref{nI}.

Satisfying the inequalities~\ref{nSe}, \ref{nC2f} and~\eqref{eq:kcond},
and requiring that the $L_r$ are integral classes, leaves only two possible
values for $k_2$ and $k_3$,  
\begin{equation}
\begin{gathered}
   k_2 = 3 , \quad k_3 = 5 , \quad \text{ giving } k = 1 \\
   \text{or} \\
   k_2 = 3 , \quad k_3 = 6 , \quad \text{ giving } k = 3
\label{eq:ksol}
\end{gathered}
\end{equation}

In general, $M$ lies in a three-dimensional subspace spanned by
\begin{equation}
   e'_4 - e'_5 , \qquad
   e'_4 - e'_6 , \qquad
   3l' - 2e'_4 - 2e'_5 - 2e'_6 - 3e'_7 . 
\label{eq:Mspan}
\end{equation}
We will restrict ourselves to the one-dimensional subspace
$M=z(e'_4-e'_5)$ for some integer $z$. (Other solutions exist with
more general $M$.) 

Finally, we are left to satisfy the second stability
condition~\ref{nSs} and the inequality~\ref{nC2fp}. It is
straightforward to show that there is a solution
\begin{equation}
   k_2 = 3 , \qquad k_3 = 6 ,
\label{eq:kfinal}
\end{equation}
together with 
\begin{equation}
   \label{eq:uxzsol}
   u = -3 , \qquad z = 1, \qquad x = 5 .
\end{equation}
For the integers $a_{r\k}$ we have
\begin{equation}
   a_{21} = a_{22} = a , \qquad
   b_{21} = b_{22} = b_{23} = b 
\label{eq:asol}
\end{equation}
for arbitrary non-negative integers $a$ and $b$. Finally, the degrees
$d_2$ and $d_3$ have the form
\begin{equation}
   d_2 = 2p , \qquad d_3 = 3q + 1 , 
\label{eq:dsol}
\end{equation}
for arbitrary integers $p$ and $q$. 

In conclusion, we have constructed a large new class of bundles on
non-simply connected Calabi-Yau manifolds
which give three-family, anomaly-free vacua with the standard
model gauge group. From equations~\eqref{eq:asol} and~\eqref{eq:dsol},
we see that rather than a single solution, we have a class of
solutions depending on four arbitrary parameters. Furthermore, other
solutions exist, with a more general class $M$
in~\eqref{eq:Lsol}. This provides flexibility for discussing other
physical properties of these models, such as nucleon decay and Yukawa
couplings.


\acknowledgments
\label{sec:ack}

R.~Donagi is supported in part by an NSF grant DMS-9802456 as well as a
UPenn Research Foundation Grant. 
B.~A.~Ovrut is supported in part by a Senior Alexander von Humboldt
Award, by the DOE under contract No. DE-AC02-76-ER-03071 and by a
University of Pennsylvania Research Foundation Grant. 
T.~Pantev is supported in part by an NSF grant DMS-9800790 and by an
Alfred P. Sloan Research Fellowship. 
D.~Waldram would like to thank Enrico Fermi Institute at The
University of Chicago and the Physics Department of The Rockefeller
University for hospitality during the completion of this work. 



\end{document}